# MULTI-AGENT BASED PROTECTION SYSTEM FOR DISTRIBUTION SYSTEM WITH DG

JIN SHANG, NENGLING TAI and QI LIU

*Department of Electrical Power Engineering, Shanghai JiaoTong University,
Shanghai,200240,China*
*factsj@sjtu.edu.cn*

This paper introduces the basic structure of multi-agent based protection system for distribution system with DGs. The entire system consists of intelligent agents and communication system. Intelligent agents can be divided into three layers, the bottom layer, the middle layer and the upper layer. The design of the agent in different layer is analyzed in detail. Communication system is the bridge of multi-agent system (MAS). The transmission mode, selective communication and other principles are discussed to improve the transmission efficiency. Finally, some evaluations are proposed, which provides the design of MAS with reference.

## 1. Background and Introduction

The environmental problems have aroused worldwide attention, and the energy crisis is getting more and more serious. With the development of renewable sources and its controlling technology, a number of distributed generations (DG) have been connected to the distribution system, which provides a new method to save the limited energy and reduce carbon dioxide emissions.

However, DGs also bring some problems. Protection coordination is one of the most important problems. There're lots of researches that analyze the problems from different aspects [1-4]. The key point is that, there is only single power supply in the original system, but DGs add new power supplies to the grid, which make it into a complex multi power supplies network. The new power supplies will significantly affect the power flow. When a fault occurs, they will provide additional current to the fault point, which will greatly affect the reliability of protection.

Since the original protection of distribution system is no longer as reliable as before, there're many improved methods proposed to solve this problem. It is very convenient to disconnect the DG unit when a fault occurs [4]. And most standards for DG also require like this (IEEE-Std.1547-2003). In this case, DG will not affect the fault current. However, for the system with large DG penetration level, it will not support the grid voltage and frequency during and immediately after the fault. Restricting the capacity and access point of DG is another method [5-7]. Then the influence of DGs is greatly reduced. But this will greatly stunt the development and application of DG. The fault current limiter (FCL) has also been popular for a while [2,8]. There're also many other methods that proposed to improve the original protection scheme to make the protection system more reliable [9-12]. However, all these methods have some limitation. They may have solved the problems in some degree, but, they're not systemic and effective enough. The protection scheme should be modified when the structure of the system changes.

The multi-agent technology was first applied to the power system in the early 1990s. And there're lots of researches after that, such as optimal power flow algorithm, power market, voltage controlling and so on. The protection system based on multi-agent was proposed not long ago. It has been studied a lot recently [13-16]. However, the application of MAS for the protection of distribution system with DGs has been rarely analyzed before. This paper presents an effective and extensible scheme. The basic structure of MAS is introduced, the design of the agents and communication system are analyzed.

## 2. Structure of MAS

The multi-agent system is composed of different intelligent agents, which cooperate with others as individual units. Any electrical element that contains necessary information can be represented as an element agent, such as the line breaker, transformer and so on. There're also integrated agents, they do not collect information from certain element directly, but collect information from other agents and draw conclusions. Fig. 1 shows the schematic diagram of a typical distribution system and the corresponding protection system.

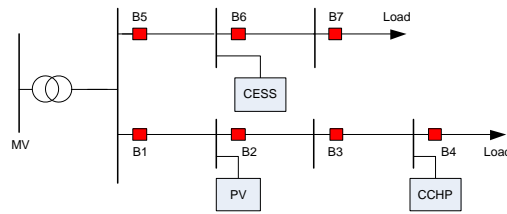

(a) Schematic diagram of distribution system

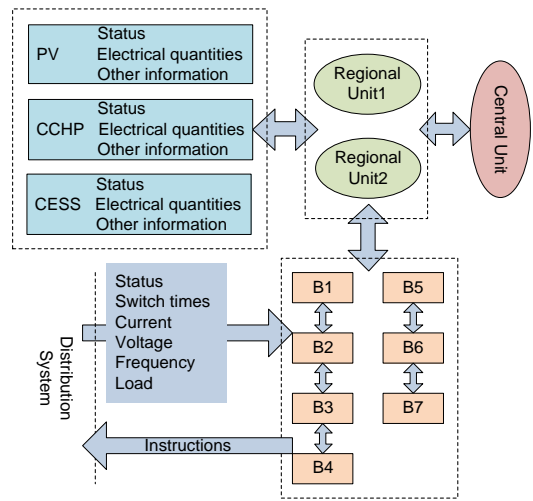

(b) Structure of corresponding MAS

Fig.1 Schematic diagram of distribution system and MAS

The distribution system in sub-graph (a) consists of two feeder lines and three energy sources, the Photovoltaic (PV), Combined Cooling Heating and Power (CCHP), and Composite Energy Storage System (CESS). In sub-graph (b), the seven orange boxes (B1 to B7) represent the branch agents. which gather information from the distribution system directly. The three blue boxes represent DG agents, which gather the real-time information of DGs. The two green ellipses represent the regional agents, which are the middle segments between terminal agents and central agent. The red ellipse represents the central agent. The arrows represent the information flow, two-way arrow means two-way flow. The arrows can be considered to make up the communication system.

According to the structure introduced above, there're mainly two aspects for the implementation of MAS. One aspect is the design of various kinds of agents. The other is the construction of communication system. The two aspects are both very important, and they influence each other at the same time. The model of agent affects the structure of communication system. Meanwhile, the restriction of communication limits the information transmission between different agents. It's of great importance to design both of them properly.

## 3. Intelligent Agents

In order to realize the function of different levels and make the structure of MAS clear, the whole MAS can be divided into three layers [17], the bottom layer, the middle layer and the upper layer, as shown in Fig.2.

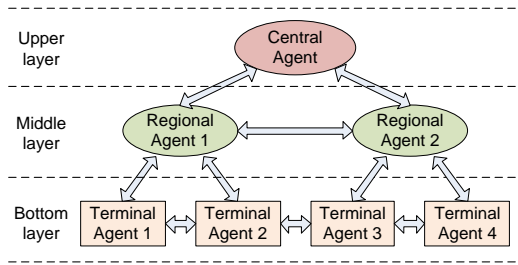

Fig.2 Hierarchy chart of agents

All the terminal agents that directly interact with the distribution system (including different DGs), make up the bottom layer. Regional agents that connect bottom layer and the upper layer make up the middle layer. The central agent gathers information of the entire system, and it forms the upper layer. Three layers play different roles. Agents are designed according to the functional requirements.

*3.1. Terminal agents*

The agents for breakers, lines, DGs and other elements in the bottom layer are classified as terminal agents. They not only gather information from the power system, but also work as actuators. The tripping, reclosing, load shedding and other instructions are all implemented through them. Since their functions are complicated, the model should be well designed.

Fig.3 shows the structure of the branch agent. The branch agent is the integration of line agent and breaker agent. Since the breaker agent provides the line with protection and the line agent provides the relay with information, the two agents are closely related with each other. Thus they're integrated to reduce the investment and make the communication between them easier.

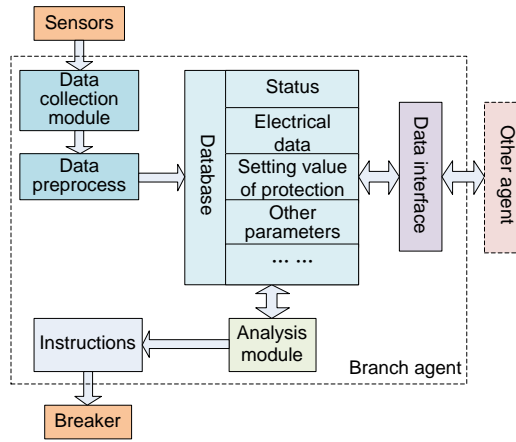

Fig.3 Structure of branch agent

As shown in Fig.3, the dashed box represents for the branch agent. Data collection module is responsible for the data acquisition from the sensors, which directly measure the electrical qualities of the power system. The data preprocess module will fulfill the data conversion and filtering, Then, the data with prescribed format will be stored in the database. The analysis module reads the data from the database and conducts the fault analysis [18]. If the breaker should act to clear the fault, the instruction will be sent out. The action will also be recorded in the database. Once the instruction is successfully executed, the agent will get the new status. The data interface is a bi-directional port connected to the database. The agent will communicate with other agents through this port.

Database is the core of the agent, all the electrical data collected from the power system are stored here, such as the status, the setting values of protection, and other parameters that related to the line and the equipment (e.g. relay, breaker). The model of the database will directly affect the efficiency of the agent. Fig.4 shows the structure of the database model.

There are more details in the practical model. Each property has its sub-properties. For example, if the "Breaker status" shows that the breaker is closed, there is a sub-property to tell whether the current stratus is normal state, reclosing or manual operation. Besides, the analog properties (current, voltage) should be stored separately according to phase a, b and c. "Set value" is concerned with the protection principle and the status of DGs. Several groups of set value are stored for different situations.

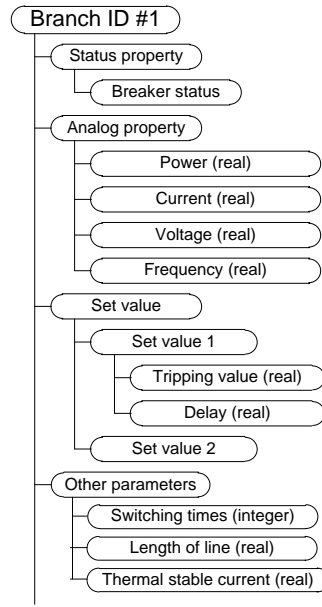

Fig.4 Structure of a database model

Besides the branch agent analyzed above, there're also transformer agent, busbar agent and so on. The design of this kind of agent is similar with the branch agent. The design of DG agent is also similar, although the internal control of DG is very complex, only the external information is the key for protection system. The external information presented by DG is a little different from that of the ordinary element, but the design idea is the same. In a word, the agent should collect information from the power system, take analysis, and communicate with other agent.

*3.2. Regional agents*

Regional agents play the role as a gateway. They do not connect with the distribution system directly, but collect information from a certain area through several terminal agents. The information is preprocessed here. The existing of regional agents can avoid the data flood. Although each regional agent covers a certain area, adjacent areas should overlap with each other to ensure the comprehensiveness and reliability of the protection system. Since the regional agents exchange information widely with others, its database, algorithm, interface and the whole structure are all very important. The structure of regional agent is shown in Fig.5.

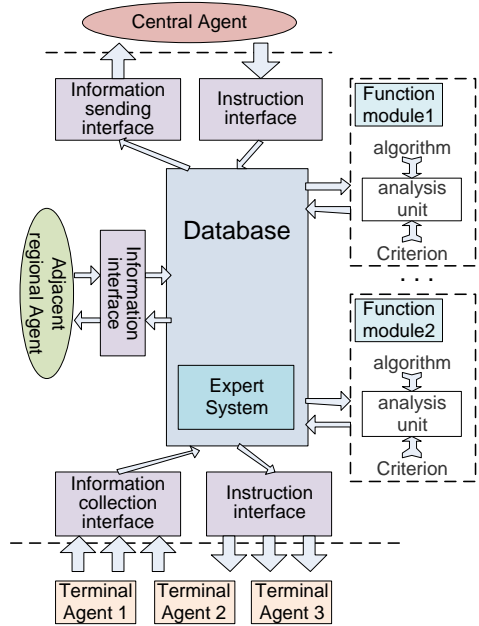

Fig.5 Structure of regional agent

When the regional agent is to be designed, the function of the agent is defined at first. Then, the information required and the instructions sent out are confirmed. The database and interfaces are designed according to this. Database lays a solid foundation for the function modules. For different function modules, it's very important to choose a proper algorithm to achieve the function quickly and accurately.

Since the influences of DGs to the protection system are complicated, MAS should be adaptive in different situations. However, the speed of online adaptive system can hardly meet the requirement of protection. But the possible running status of DGs in a certain distribution system are fixed, which provides possibility to build an offline adaptive system. An expert system is built to support the offline adaptive system. When the running status of the network changes, the expert system will provide appropriate protection knowledge, including protection configuration, coordination, setting values and so on. The knowledge as well as the real-time information will be read by the function module. Then, the new protection scheme can be figured out and instructions will be sent to the terminal agents.

*3.3. Central agents*

Since MAS has good abilities of information interaction and controlling [19-20], besides working as protection system, it is also responsible for the optimal operation of the entire system.

Some kinds of DG have a random power output, such as PV wind turbine and CCHP. Fuel cell and CESS are flexible units, which can be adjusted as source or load. Since the central agent can obtain the information of the entire system, it's able to arrange the running of each DG to achieve the optimal and stable running of distribution system. DG agents will implement the instructions. The concrete implementations are carried out by the controlling module inside of DG, which is not concerned by MAS.

Once the voltage or frequency of the distribution system is abnormal, the central agent can make adjustments by controlling the active and reactive power of DGs, transferring load or even shedding load to keep the system stable.

The goal of central agent involves the entire distribution system. The information it requires mainly comes from the regional agents. Naturally, only the conclusive and necessary messages after preprocessing are sent here. Since the controlling target of central agent is complex, the algorithm is much more complicated. Different algorithms are applied under different situations. Using appropriate algorithms for different application is of great significance.

## 4. Communication System

Communication system is the bridge of MAS. The transmission mode, efficiency, synchronization, cost and information securities are important aspects to evaluate the system. They not only challenge the communication technology, but also put forward higher demands on the configuration scheme of MAS.

### *4.1. Transmission mode*

Using proper transmission mode will significantly improve the performance of MAS. There're three communication modes that commonly used, direct communication, radio communication, and blackboard system. Direct communication is applied for the situation, in which both sides know each other in advance. Radio communication enables one agent to broadcast information to all the members in the same group, even without knowing the address. Blackboard system is a centralized controlling system, in which all the group members upload their information to a public area and the others can share the information as they need. The three modes are applied in different situations according to their characteristics.

Terminal agents directly contact with the power system. They demand little of information analysis. So, it's unnecessary for them to communicate widely with others. They only communicate with adjacent agents, which is fixed and regular. It's better to use direct communication. Terminal agents also send information to regional agent. Since each regional agent covers a certain area, it will be determined which few terminal agents are in this area. In turn, these terminal agents will clearly know their communication object. So, it's also suitable to use direct communication. There're lots of terminal agents in bottom layer, direct communication reduces the cost and make the communication more safe and effective.

A regional agent is the core of the certain area. Besides receiving information from the terminal agents, it also sends instructions to them. Since there is more than one terminal agent in the group, the instructions may be sent to any one of them. It's better to use radio communication, then, all the agents in this group can receive the information and the instruction will be executed by the corresponded one after further confirmation. Regional agents also communicate with the other agents in the same layer. Blackboard system is applied in this situation. All the regional agents put their information in the public area. Others can easily get what they need for the further analysis. Because the public area is accessible for all the regional agents, it requires strong fault tolerance. When the messages uploaded by different agents are in conflict, only the right one or the one with higher priority will be recorded. The conclusions made by the regional agents should be sent to the central agent, which is similar to the information transmission from terminal agent to regional agent. The direct communication is suitable in this case.

The central agent gets information from all the regional agents for global optimization. The results and instructions will then be sent to the regional agents by radio communication.

The communication modes between different objects in MAS are listed in Tab.1.

Tab. 1 Communication mode of MAS

| Sender / Receiver | Terminal agent | Regional agent | Central agent |
|---|---|---|---|
| Terminal agent | Direct | Radio | —— |
| Regional agent | Direct | Blackboard | Radio |
| Central agent | —— | Direct | —— |

### *4.2. Selective communication*

In order to save costs and increase speed, the communication system should be effective. When the agent communicates with others, the objects should be selective. In the bottom layer, terminal agent communicates with adjacent agents. As shown in Fig.1, line B7 is far from line B4, whether breaker B7 should act or not almost has nothing to do with line B4. So it's unnecessary to connect agent B7 with agent B4. On the contrary, line B1 and B2 are adjacent, which make them closely related. It's very important to acquire the information of the adjacent line to make fault analysis. So the two agents are connected to share the information.

The information transferred should also be selective. For the communication between terminal agents, the breaker status and current information is useful for the others, while the voltage and switch times are useless. It's of more importance to make it selective when the terminal agents send information to the regional agents and the regional agents send information to the central agent. Regional agents and central agent get a large amount of information from more than one agent, it must be ensured that only the information necessary for the further analysis are received. What's more, central agent is more likely to receive the analysis results from the regional agents to achieve the rapid analysis of optimization operation.

*4.3. Principles*

There're some other principles that should be followed to improve the efficiency and reduce the cost of the communication system.

Firstly, the task of local agent should be arranged reasonably. Since all the agents are intelligent, they should carry out analysis as much as possible within their capability. This will reduce the information transmission from the source.

Secondly, it's better to transport status information instead of analog information [21]. For example, directional overcurrent protection needs current direction as criterion, while low voltage start overcurrent protection needs the value of voltage. Direction can be easily represented by a status property, which takes up only 1 byte. But the value of voltage takes up more memory and it's more demanding on the synchronization of information. The transmission of status information is easier and more reliable.

The transmission distance should be as short as possible. A terminal agent should send information to the nearest regional agent rather than the other one. Meanwhile, terminal agents do not send information directly to central agent. The information is sent to regional agent and preprocessed to get simplified conclusions. The conclusions will then be sent to the central agent.

5. Conclusions

The design of MAS can be more flexible on the premise of reliability. An excellent MAS implementation should at least meet four requirements as following:

First of all, it must satisfy the reliability, sensitivity, rapidity, selectivity requirements of protection and achieve the optimal controlling of the distribution system.

Second, the modules of MAS should be standardized to ensure the compatibility of different systems and good scalability to the expansion of grid.

Third, the construction and running of MAS should be economical.

Fourth, the information security should be guaranteed.

The four requirements define the optimal goals and provide the design of MAS with reference. Generally, the future protection of distribution system should be intelligent based on the multi-agent technology and the communication network. With the development of DGs and its controlling technology, a communication network for DGs should be built to achieve the optimal running of power system. This will also lay foundation for the multi-agent based protection system and makes it an inexorable trend.

References

1. Yuan Chao, Wu Gang, Protection technology for distributed generation systems, *Power System Protection and Control*, **37**(2009) 99-105.
2. K. Kauhaniemi, L. Kumnpulainen, Impact of distributed generation on the protection of distribution networks, *in Proc. et lEE Int. Conf.on Developments in Power System Protection*. Vol. 1(2004), pp. 315 - 318.
3. Vandevelde, L, Melkebeek, J, Protection of Power Distribution Network with High Penetration Level of

Distributed Generation, *Fifth FTW PhD Symposium, Faculty of Engineering, Ghent University*, 2004, paper No. 024.
4. Morren, J., Haan, S.W.H. Impact of distributed generation units with power electronic converters on distribution network protection, *IET 9th International Conference on Power Energy, & Industry Applications, Glasgow, UK*, 2008
5. Wang Jianghai, Tai Nengling. Penetration level permission of DG in distributed network considering relay protection, *Proceedings of the CSEE*, **30**(2010) 37-43.
6. Feng Xike, Tai Nengling. Research on the impact of DG capacity on the distribution network current protection and countermeasure, *Power System Protection and Control*, **38**(2010) 156-165.
7. Lei Jinyong，Huang Wei，Xia Xiang．Penetration level calculation with considerations of phase-to-phase short circuit fault, *Automation of Electric Power Systems*，**32**(2008) 82-86
8. YE Lin, LIN Liang-zhen. Superconducting Fault Current Limiter Applications in Electric Power Systems, *Proceedings of the CSEE*, **20**(2000) 1-5
9. Lin Xia, Lu Yuping, Wang Lianhe. New Current Protection Scheme Considering Distributed Generation Impact, *Automation of Electric Power System*, **32**(2008) 50-56.
10. Zhang Yanxia, Dai Fengxian. New Schemes of Feeder Protection for Distribution Networks Including Distributed Generation, *Automation of Electric Power System*, **33**(2009) 71-74.
11. Li Yongli, Jin Qiang, Li Botong. Application of inverse-time overcurrent protection based on low voltage acceleration in Micro-Grid, *Journal of Tianjin University*, **44**(2011) 955-960.
12. Sun Jingliao, Li Yongli. Study on Adaptive Current Instantaneous Trip Protection Scheme for Distribution Network with Inverter Interfaced DG, *Automation of Electric Power System*, **33**(2009) 71-76.
13. Renan Giovanini, Kenneth Hopkinson, Denis V. Coury, and James S. Thorp, A Primary and Backup Cooperative Protection System Based on Wide Area Agents, *IEEE Transactions on Power Delivery*, **21**(2006) 1222-1230
14. Wang Huifang, He Benteng, Shi Hongyu, Applications of multi-Agent technology in protection system, *Electric Power Automation Equipment*. **27**(2007) 102-106.
15. COURY D V, THORP J S, HOPKINSIN K M. An agent- based current differential relay for use with a utility Intranet. *IEEE Transactions on Power Delivery*, **17**(2002) 47-53.
16. HOSSACK J A, MENAL J, STEPHEN D J, et al. A multi-agent architecture for protection engineering diagnostic assistance. *IEEE Transactions on Power Systems*, **18**(2003) 639-647.
17. Xu Tianqi, Yin Xianggen, Analysis on functionality and feasible structure of wide area protection system, *Power System Protection and Control*, **37**(2009) 93-97.
18. K.K. Li, W.L. Chan, Xiangjun Zeng and Xianzhong Duan, Agent-based self-healing protection system, *IEEE Transactions on Power Delivery*, **21**(2006) 610-618
19. I. Zabet and M. Montazeri, Implementing Cooperative Agent-based Protection and Outage anagement System for Power Distribution Network Control, *4th International Power Engineering and Optimization Conf*, Shah Alam, Selangor, MALAYSIA, 2010.
20. H. N. Aung, A. M. Khambadkone, D. Srinivasan, and T. Logenthiran, Agent-based Intelligent Control for Real-time Operation of a Microgrid, *2010 Joint International Conf. Power Electronics, Drives and Energy Systems (PEDES) & 2010 Power* India, New Delhi, India, 2010
21. Sun Hui, Liu Qianjin, Power Distribution Network Protection and Control Using Mas, *Proceedings of the CSU-EPSA*. **23**(2011) 135-141.